\documentclass[showpacs,preprintnumbers, amsmath,amssymb,superscriptaddress,prep
rint]{revtex4}
\usepackage{graphicx}
\draft
\begin{document}
\newcommand {\be}{\begin{equation}}
\newcommand {\ee}{\end{equation}}
\newcommand {\bea}{\begin{eqnarray}}
\newcommand {\eea}{\end{eqnarray}}
\newcommand {\nn}{\nonumber}


\title{Quasiparticle spectrum of the hybrid s+g-wave superconductors
YNi$_2$B$_2$C and LuNi$_2$B$_2$C
}

\author{Kazumi Maki}
\address{Department of Physics and Astronomy, University of Southern
California, Los Angeles, CA 90089-0484}  

\author{Hyekyung Won}
\address{Department of Physics, Hallym University, Chunchon 200-702,
South Korea
}

\author{Stephan Haas}
\address{Department of Physics and Astronomy, University of Southern
California, Los Angeles, CA 90089-0484}  

\date{\today}
\textrm{\textit{}}

\begin{abstract}
Recent experiments on single crystals of YNi$_2$B$_2$C have revealed the
presence of point nodes in the superconducting energy gap $\Delta( {\bf k})$
at $\hat{k}$ = (1,0,0), (0,1,0), (-1,0,0), and (0,-1,0).  In this paper we 
investigate the effects of impurity scattering on the quasiparticle spectrum
in the vortex state of s+g-wave superconductors, which is found to be 
strongly modified in the presence of disorder. In particular, a gap in 
the quasiparticle energy spectrum is found to open even
for infinitesimal impurity scattering,
giving rise to exponentially activated thermodynamic response functions,
such as the
specific heat, the spin susceptibility, 
the superfluid density, and 
the nuclear spin lattice relaxation.
Predictions derived from 
this study can be verified by measurements of the angular dependent 
magnetospecific heat and the magnetothermal conductivity. 
  
\end{abstract}
\maketitle

\noindent{\bf \it 1. Introduction}

The recently discovered superconductivity in the borocarbides 
YNi$_2$B$_2$C and LuNi$_2$B$_2$C has attracted much attention due to 
the rather unusual apparent structure of their gap function $\Delta( {\bf k})$.
\cite{izawa1,park} On the one hand, the presence of a substantial 
s-wave component has been observed by thermodynamic measurements of samples
in which Ni was substituted by small amounts of Pt, causing the opening
of an energy gap.\cite{nohara1,borkowski} On the other hand, there is
also mounting experimental evidence for the presence of nodes
in the superconducting order parameter.\cite{nohara2,zhang,izawa2}
A consistent way to account for these seemingly contradictory 
observations is to consider the hybrid s+g-wave order parameter
\cite{maki1,thalmeier}, given by
\bea
\Delta({\bf k}) = \Delta \left( 1 - \sin^4 (\theta ) \cos (4 \phi )
\right) /2 ,
\eea
where the angles $\theta$ and $\phi$ indicate the direction of the 
quasiparticle wave vector ${\bf k}$. A graphical represenation of 
$\Delta({\bf k}) $ is given in Fig. 1(a). This order parameter has 
recently been shown to account for the Raman spectra measured in 
YNi$_2$B$_2$C and LuNi$_2$B$_2$C.\cite{yang,jang} Furthermore, it
was observed that the corresponding quasiparticle spectrum is very sensitive
to impurity scattering. In particular, the thermal conductivity of
Y(Ni$_{1-x}$Pt$_x$)$_2$B$_2$C for $x$=0.05 in the vortex state at 
T=0.8K does not exhibit any angular dependence, contrary to the clear
cusp structure observed in the pure system.\cite{kamata}

\begin{figure}[h]
\includegraphics[width=5cm]{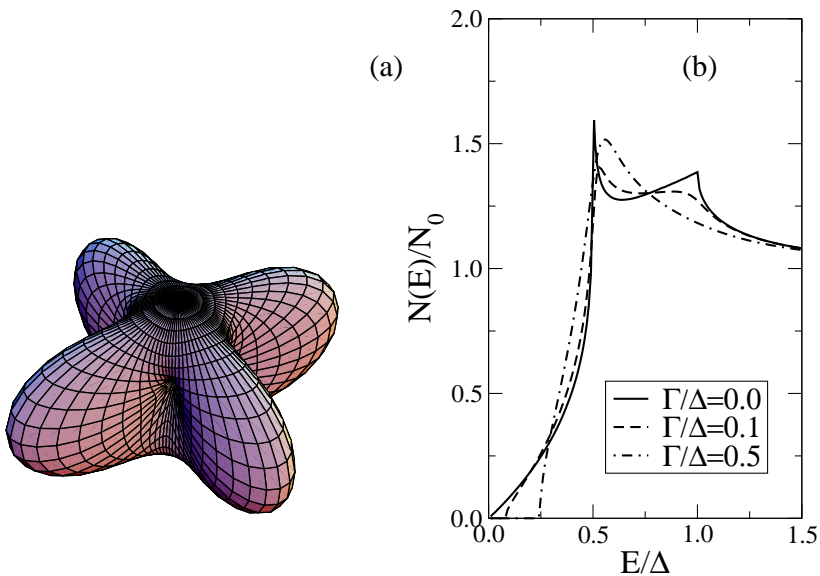}
\includegraphics[width=6cm]{fig1b.eps}
\caption {(a) s+g-wave order parameter with four point nodes,
(b) quasiparticle density of states for various impurity scattering rates.
}
\end{figure}

Theoretical studies of nonmagnetic impurity scattering in anisotropic 
superconductors indicate that this type of disorder has much 
more profound effects on the hybrid s+g-wave superconductors with 
point nodes than on other unconventional superconductors with line nodes. 
\cite{won1,yuan}
In particular, it was observed that the unitary limit 
and the weak-scattering Born limit give practically
the same results over a 
wide range of parameters. Also,
contrary to superconductors with line nodes,  
no low-frequency resonance is found for the strong-scattering case. 
Even more surprisingly, a disorder-induced quasiparticle energy gap 
$\omega_g$ opens up 
even for infinitesimal scattering rates $\Gamma$. This dependence is 
well approximated by $\omega_g = \Gamma /\left( 1 + 2 \Gamma /\Delta
\right)$.\cite{won1,yuan}
Typical quasiparticle densities of states for the pure s+g-wave
system and in the presence 
of impurity scattering are shown in Fig. 1(b).

The opening of a disorder-induced quasiparticle energy gap offers 
a natural explanation for the reported exponentially 
activated temperature dependence of the thermal conductivity and 
the specific heat,
\bea
\frac{C_S}{T} , \frac{\kappa_{zz}}{T} \propto 
\left(\frac{\omega_g}{T}\right)^{3/2} \exp (-\omega_g /T).
\eea
For the same reason, no universal heat conduction is expected for
these systems as opposed to superconductors with line nodes.\cite{lee,sun}   
Because these effects of impurity scattering are rather singular compared with 
other nodal superconductors, it is necessary to reexamine how transport
properties are affected by disorder.\cite{maki1,thalmeier} One finds 
that in the superclean limit, i.e. $\Gamma \ll \sqrt{v_a v_c e H}$
where $v_a$ and $v_c$ are the Fermi velocities along the crystallographic
a- and c-directions, the 
thermal conductivity $\kappa_{zz}$ has a similar angular  
dependence as in the pure system.\cite{maki1} However, the opening of the 
energy gap due to the impurity scattering drastically reduces
nodal excitations. For $\Gamma/\sqrt{v_a v_c e H} > \pi/\sqrt{2}$
practically all of the nodal excitations are eliminated, and therefore
the angular dependence of the thermal conductivity is suppressed, as it
was recently observed by Kamata {\it et al.}.\cite{kamata} 

The impurity scattering rate $\Gamma$ can be accurately estimated in
terms of its effect on the superconducting transition temperature,
\cite{won1,yuan}
\bea
-\ln \left( \frac{T_c}{T_{c0}} \right) = \frac{\langle f^2 \rangle}
{1 + \langle f^2 \rangle } \left[ \psi \left( \frac{1}{2} +
\frac{\Gamma}{2 \pi T} \right) - \psi \left( \frac{1}{2} \right) \right],
\eea
where $\psi (z) $ is the di-gamma function, $f \equiv \sin^4 (\theta ) 
\cos (4 \phi )$ denotes the angular dependence of the s+g-wave gap 
function, $\langle f^2 \rangle \simeq 0.203$ results from angular 
averaging, and $T_{c0}$ is the critical superconducting transition 
temperature of the pure system. For the 5\% Pt doped YNi$_2$B$_2$C,
the reduced critical temperature $T_c$ = 13.5K therefore indicates a 
scattering rate $\Gamma$ = 28.3K, where $T_{c0}$= 15.5 K was used.
This implies that for the doping level $x \sim 0.05$ the corresponding 
disorder induced energy gap is $\omega_g \sim $ 10K. Since this is 
already a rather strong effect, it may be necessary to measure single
crystals with $x <$ 0.01, in order to observe
the associated disappearance of the angular dependence in the 
thermal conductivity. 

\noindent{\bf \it 2. Quasiparticle Density of States in the Vortex State} 

The quasiparticle density of states and the quasiparticle scattering
rate are obtained from the integral\cite{won2}
\bea
I = \left( x + i C_0 \right) \langle \left( \left( x + i C_0 \right)^2
- \left( (\Gamma / \Delta) + (1 - f)/2 \right)^2 \right)^{-1/2} \rangle,
\eea
where $x \equiv \langle | {\bf v} \cdot {\bf q} | \rangle / \Delta$, and
$ {\bf v} \cdot {\bf q} $ is the Doppler shift due to the supercurrents
circulating around the vortices.\cite{volovik,won2} 
The terms $(\Gamma / \Delta )$ and $C_0$ in Eq. 4 arise from the renormalization
of the order parameter and of the frequency due to the presence of
impurity scattering,
\bea
\tilde{\Delta } |_{\omega = 0} & = & \Gamma,\\
\tilde{\omega} |_{\omega = 0} & = & i C_0 \Delta.
\eea

For the gap symmetry
of $\Delta ({\bf k}) $ given in Eq. 1, and in a magnetic field ${\bf H}$ 
directed along $(\theta , \phi )$ with the polar axis along the crystal
c-direction, one obtains
\bea
x \simeq \frac{ \sqrt{v_a v_c e H}}{2 \Delta} \left(
\sqrt{1 - \sin^2 \theta \cos^2 \phi} + \sqrt{1 - \sin^2 \theta \sin^2 \phi} 
\right) .
\eea
For small impurity scattering $\Gamma / \Delta \ll 1 $, the angular
averages Eq. 4 can be 
performed analytically, giving
\bea
I \simeq \frac{1}{2} \left( x + i C_0 \right)\left( \cosh^{-1} \left(
\frac{1}{x} \right) + \cos^{-1} y - \frac{i C_0}{x} \left( 1 - \frac{y}
{\sqrt{1 - y^2}} \right) \right) 
,
\eea
where $y \equiv \Gamma / x \Delta $. 
Within this limit, one obtains closed expressions for
the quasiparticle density of states $g( {\bf H}, \Gamma )$ and for the 
quasiparticle damping constant $C_0 \Delta$:
\bea
g( {\bf H}, \Gamma )  \equiv  \frac{N(E) }{N_0} &=& 
\frac{x}{2} \cos^{-1}y \theta (1 - y) + {\cal O}(C_0),\\
C_0 \Delta
& = & \frac{\Gamma x}{2 } \left[ \ln \left( \frac{2}{x} \right) 
+ {\cal O}(C_0) \right].
\eea
Consequently, the low-temperature specific heat, the spin susceptibility,
the superfluid density in the a-b plane, and 
the nuclear spin lattice relaxation rate are given by
\bea
\frac{C_S}{\gamma_N T} & = & \frac{\chi_S}{\chi_N} = g( {\bf H}, \Gamma ),\\
\frac{\rho_{Sab} ({\bf H})}{\rho_{Sab} (0)} & = & 1 - \frac{3}{2} 
g( {\bf H}, \Gamma ),\\
\frac{T_1^{-1} }{T_{1N}^{-1}} & = & g^2( {\bf H}, \Gamma ).
\eea
Note that in contrast to $\rho_{Sab} ({\bf H})$ the superfluid density
along the c-direction $\rho_{Sc} ({\bf H})$ is not 
linear in $g( {\bf H}, \Gamma )$. 

\begin{figure}[h]
\includegraphics[width=9cm]{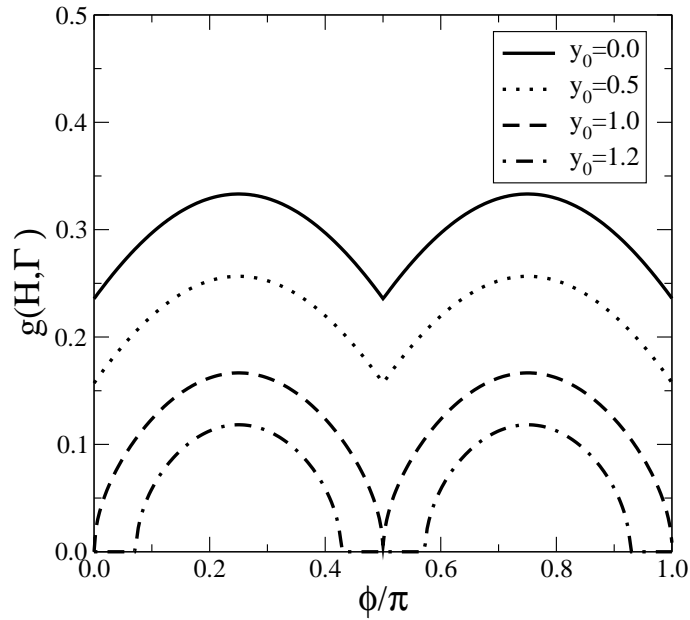}
\caption {Dependence of the quasiparticle density of states
on the azimuthal angle $\phi$ for various impurity scattering 
strengths. The polar angle is fixed to $\theta = \pi /2$. \label{fig2}}
\end{figure}

In Fig. 2, the dependence of the quasiparticle density of states 
$g( {\bf H}, \Gamma )$ on the azimuthal angle $\phi$ is shown for a fixed
polar angle
$\theta = \pi / 2$, and for $y_0$ = 0.0, 0.4, 1.0, and 1.2, where
$y_0 \equiv (2 \Gamma)/(\pi \sqrt{v_a v_c e H} )$ is a measure
of the impurity scattering strength. From Eq. 9 it follows
that $g( {\bf H}, \Gamma ) = 0$ for $y_0 > \sqrt{2}$. Furthermore, it is 
observed in Fig. 2 that the $\phi$-independent part of $g( {\bf H}, \Gamma )$
gets cut off for increasing $\Gamma$.  For $y_0 >$ 1 the constant part
is completely eliminated, and only small islands survive for $1 < y_0 <
\sqrt{2}$. This is a signature of a ${\bf k}$-independent 
energy gap $\omega_g$ that opens up due to impurity scattering. Thermodynamic
response functions, such as the specific heat, the spin susceptibility,
the superfluid density,
and the nuclear spin lattice relaxation rate provide a direct way 
to study this disorder-induced phenomenon. 

\noindent{\bf \it 3. Angular Dependent Thermal Conductivity}

Following the procedure given in Ref. \cite{won2}, the diagonal
components of the thermal 
conductivity tensor are given by
\bea
\frac{\kappa_{zz}}{\kappa_n} & = & \frac{x}{2 \ln (2/x)} 
\left( (1-y^2)^{3/2} - \frac{3 y }{2} (\cos^{-1} y - y \sqrt{1 - y^2} )
\right) \theta(1 - y),\\
\frac{\kappa_{xx}}{\kappa_n} & = & \frac{3}{2 \ln (2/x)} \left( \frac{x'}{x}
\right)^2
\left( \cos^{-1} y' - y' \sqrt{1 - y'^2} 
\right) \theta(1 - y'),
\eea
where $x' \equiv \sqrt{v_a v_c e H} \sqrt{1 - \sin^2 \theta \cos^2 \phi }/
2 \Delta$ and 
$y' \equiv y_0 / |\sin \phi | $.

\vspace{1.0cm}
\begin{figure}[h]
\includegraphics[width=9cm]{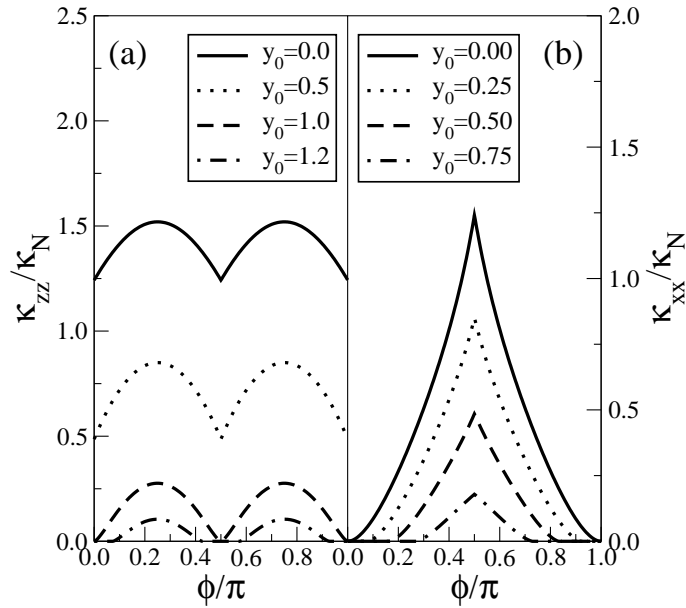}
\caption {Angular dependent thermal conductivity of a 
s+g-wave superconductor in a magnetic
field. (a) zz-component, and (b) xx-component of the 
thermal conductivity tensor
for various impurity scattering 
strengths. The polar angle is
fixed to $\theta = \pi /2$. \label{fig3}}
\end{figure}

The angular dependence of $\kappa_{zz}$ and $\kappa_{xx}$ is shown 
in Figs. 3(a) and (b). In the limit 
$y_0 \rightarrow 0$ Eq. 14 has a similar angular dependence
as given in Ref. \cite{thalmeier}, with the exception of a weak 
$\ln x$ correction in the denominator. Moreover the dependence on the 
impurity scattering strength $\Gamma$ is similar to the 
one for the quasiparticle
density of states described in the previous section.
On the other hand, the structure of the angular dependence of $\kappa_{xx} $ 
described in Eq. 15 is 
dominated by an inverted cusp centered at $\phi = \pi/2$. 
Furthermore, we note that for small applied magnetic fields $\kappa_{zz}$ 
increases like $\sqrt{H}$, whereas $\kappa_{xx}$ remains independent of $H$. 
The reason for this is that the current operator $\hat{Q}_z$ vanishes at the 
nodal points in the $k_x-k_y$ plane, whereas $\hat{Q}_x$  is non-zero in
the nodal plane. This peculiar effect should be explored experimentally.

We observe that there are a number of similarities of the 
order parameter of these hybrid s+g-wave compounds
with the newly discovered heavy fermion superconductor
PrOs$_4$Sb$_{12}$. In particular, the singlet
s+g-wave gap function can describe 
consistently the angular dependent thermal conductivity data for this
compound.
\cite{izawa3,maki2}
However, the presence of remnant magnetization observed in muon spin
rotation measurements is suggestive of spin triplet pairing.\cite{aoki}
More recently, Chia {\it et al.} have reported a surprisingly isotropic
superfluid density in the B-phase of PrOs$_4$Sb$_12$.\cite{chia} It
appears that none of the currently proposed models for superconductivity 
in this material match this experimental data.\cite{ichioka,miyake,goryo}
Clearly, more experimental measurements on this compound are necessary to
settle this issue.

\noindent{\bf \it 4. Concluding Remarks}

We have seen that impurity scattering has surprisingly profound effects
on s+g-wave superconductors such as YNi$_2$B$_2$C and LuNi$_2$B$_2$C.  
In particular measurements of the
magnetothermal conductivity provide a unique window to observe the
disorder induced opening of a gap in the quasiparticle energy through their
angular dependence. Furthermore, there is evidence that the skuttertide 
PrOs$_4$Sb$_12$ may also have a hybrid s+g-wave superconducting order 
parameter. The discovery of these point-node compounds opens up new
opportunities to explore unconventional superconductivity.

We thank P. Thalmeier, K. Kamata, K. Izawa, and Y. Matsuda
for useful discussions on borocarbide superconductors, 
and acknowledge financial support by the National Science Foundation,
Grant No. DMR-0089882.

\end{document}